\documentclass{article}
\usepackage{spconf,amsmath,graphicx,hyperref}

\usepackage{fix-cm}
\usepackage{amsmath}   % for \text in math
\usepackage{booktabs}  % for \toprule, \midrule, \bottomrule
\usepackage{graphicx}  % for \resizebox
\usepackage{multirow}
\usepackage{subcaption}
\usepackage{multirow}
\usepackage{enumitem}
\usepackage{booktabs}
\usepackage{xcolor}
\usepackage[fontsize=9pt]{fontsize}

\usepackage[numbers,sort&compress]{natbib}
\let\OLDthebibliography\thebibliography
\renewcommand\thebibliography[1]{
  \OLDthebibliography{#1}
  \setlength{\parskip}{4pt} % Minimal paragraph spacing
  \setlength{\itemsep}{-0.5ex} % Slightly negative item spacing
}

\title{Understanding Frechet Speech Distance for Synthetic Speech Quality Evaluation}

\name{June-Woo Kim$^{*1}$$\thanks{\hspace{-1em}$^{*}$ work done while intern at Amazon Science \, $ ^{\dagger}$corresponding author. \\ This work was partially supported by the National Research Foundation of Korea (NRF) grant funded by the Korea government (MSIT) (Grant no. RS-2025-16066662), and partially by the Bio Industry Technology Development Program funded by the Ministry of Trade, Industry \& Energy (MOTIE, Korea) (Project Number: RS-2024-00431485).}$, Dhruv Agarwal$^{\, \dagger2}$, Federica Cerina$^{2}$}
\address{$^{1}$Gwangju Institute of Science and Technology, Republic of Korea, $^{2}$Amazon Science, USA}

\begin{document}

\maketitle

\begin{abstract}

Objective evaluation of synthetic speech quality remains a critical challenge. Human listening tests are the gold standard, but costly and impractical at scale. 
Fréchet Distance has emerged as a promising alternative, yet its reliability depends heavily on the choice of embeddings and experimental settings.
In this work, we comprehensively evaluate Fréchet Speech Distance (FSD) and its variant Speech Maximum Mean Discrepancy (SMMD) under varied embeddings and conditions.
We further incorporate human listening evaluations alongside TTS intelligibility and synthetic-trained ASR WER to validate the perceptual relevance of these metrics. 
Our findings show that WavLM Base+ features yield the most stable alignment with human ratings. While FSD and SMMD cannot fully replace subjective evaluation, we show that they can serve as complementary, cost-efficient, and reproducible measures, particularly useful when large-scale or direct listening assessments are infeasible. Code is available at \textcolor{blue}{\href{https://github.com/kaen2891/FrechetSpeechDistance}{https://github.com/kaen2891/FrechetSpeechDistance}}.

%which is a prevalent metric in generative AI, can offer a potential alternative for synthetic speech evaluation. However, the optimal embeddings and datasets for accurate Fréchet Speech Distance (FSD) computation remain unexplored, and there is currently no consensus within the TTS domain regarding their effectiveness. This paper presents a comprehensive evaluation of FSD for synthetic speech quality, comparing it with other relevant metrics. Our findings demonstrate that self-supervised WavLM Base+ features yield stable FSD scores while exhibiting sample efficiency, noise robustness and alignment with synthetic audio trained ASR WER, suggesting consistency with other objective metrics.

\end{abstract}


% -------------------------------------------------------------------------

\section{Introduction}
\label{sec:intro}
Recent advancements in text-to-speech (TTS) models have shown promising results in generating high-quality synthetic samples% in zero-shot multi-speaker settings
~\cite{casanova2024xtts, le2024voicebox, casanova2022yourtts}. 
To measure the intelligibility of synthetic samples, Word Error Rate (WER) metric can be used, i.e., the lower WER implies that the synthetic samples are more intelligible and faithful to the input text prompt.
However, WER metric does not always reflect superior overall TTS quality; If the synthetic samples are generated by diverse speakers including unseen speakers or noisy acoustic prompts, the WER may increase because the ASR model struggles to understand the varied audio styles, and the higher WER in this context does not signify that the synthetic samples are of inferior quality~\cite{le2024voicebox}. 

Human listening evaluation often measured by Mean Opinion Score (MOS), remains the gold standard for assessing synthetic speech quality. Nevertheless, MOS is inherently subjective, prone to individual biases and potential misinterpretations~\cite{streijl2016mean}. Furthermore, MOS is a costly resource-intensive method and limits objective comparisons, thereby hindering reproducibility in relation to other TTS studies. Moreover, MOS/TTS intelligibility do not quantitatively measure the similarity of synthetic and real speech distributions. Methods such as comparing WER on real audio, between ASR models trained with synthetic samples and those trained solely on real audio have been proposed~\cite{le2024voicebox}. However, these approaches rely on training ASR models, which can be costly and subject to challenges like hyperparameter optimization.

%In that respect, human listening evaluation using the Mean Opinion Score (MOS) is considered the gold standard, as humans can effectively assess how natural and similar the synthetic samples are to the given acoustic prompts. However, MOS is inherently subjective, relying on individual listeners' perceptions, which can lead to potential misinterpretation~\cite{streijl2016mean}. Besides, its major drawback is that it involves subjective assessment, limiting the ability to make fair comparisons with other studies and making it a costly solution. Moreover, MOS/TTS intelligibility do not quantitatively measure the similarity of synthetic and real speech distributions. Methods such as comparing WER on real audio between ASR models trained with synthetic samples and those trained solely on real audio have been proposed ~\cite{le2024voicebox}. However, these approaches rely on training ASR models, which can be costly and subject to challenges like hyperparameter optimization.

In this context, Fréchet Distance (FD) can be employed to evaluate both the quality and similarity of generated samples relative to a reference set~\cite{dowson1982frechet}, and it has been already utilized across generative domains, including image (FID)~\cite{heusel2017gans}, music (FAD)~\cite{48813, gui2024adapting, tailleur2024correlation}, and speech (FSD)~\cite{le2024voicebox, Bińkowski2020High}. 
Nevertheless, there is still no consensus in the TTS domain on the preferred embeddings and datasets for effectively assessing synthetic speech quality. 
%Despite the success of FD in other domains, there is still no consensus in the TTS domain on the preferred embeddings and datasets for effectively assessing synthetic speech quality. 
Since FD is a distance based metric, its results are highly dependent on the specific experimental settings~\cite{bińkowski2018demystifying}. These factors can significantly impact FSD scores, making it challenging to compare reported results and potentially leading to confusion among readers.
%In this context, Fréchet distance can be employed to evaluate both the quality and similarity of generated samples relative to a reference set~\cite{dowson1982frechet}, and it has been already utilized in various generative domains, including image (FID)~\cite{heusel2017gans}, music enhancement and generation (FAD)~\cite{48813, gui2024adapting, tailleur2024correlation}, and synthetic speech (FSD)~\cite{le2024voicebox, Bińkowski2020High} analysis, respectively. Although FID and FAD are well-studied, with established protocols and various embeddings for enhancing measurement~\cite{chong2020effectively, gui2024adapting}, there is still no consensus in the TTS domain on the preferred embeddings, reference sets, and target sets for effectively assessing synthetic speech quality. Since Fréchet distance is a distance based metric, its results are highly dependent on the specific experimental settings~\cite{bińkowski2018demystifying}. These can significantly impact FSD scores, making it challenging to compare reported results and potentially leading to confusion among readers. 

%In this paper, we define speech quality in terms of its effectiveness in improving ASR performance when synthetic samples are used for training. 
In this paper, we systematically investigate how various speech embeddings, noise, and synthetic samples affect the FSD when the reference set is fixed. Our contributions are: % summarized as follows:

\begin{itemize}[itemsep=0pt, topsep=0pt, parsep=0pt, partopsep=0pt, leftmargin=*]
    \item We explore FSD using embeddings from recent self-supervised speech models~\cite{baevski2020wav2vec, hsu2021hubert, chen2022wavlm}, Whisper~\cite{radford2023robust} and ECAPA-TDNN~\cite{desplanques2020ecapa}.

    % Audio Quality Margin
    \item We introduce Speech Maximum Mean Discrepancy (SMMD) as a complementary measure to FSD.

    %class imbalance
    \item We conduct human MOS tests to validate the perceptual relevance of FSD and SMMD.

    %angular classifier
    \item We demonstrate FSD's robustness to sample bias and show that WavLM Base+ embeddings yield stable and reliable benchmarking of synthetic speech quality.

    %\item Code is available at \url{www.github}.

\end{itemize}

\section{Background}
\label{sec:Pre}
%Fréchet (Wasserstein-2) distance~\cite{dowson1982frechet} metric is widely used to measure the similarity between real and synthetic sets, especially in the domains of generative AI. 
The Fréchet (Wasserstein-2) distance~\cite{dowson1982frechet} is defined as:
\begin{align}
\mathbf{F}(\mathcal{N}_r, \mathcal{N}_g) = ||\mu_r, \mu_g||^2 + tr(\Sigma_r, \Sigma_g - 2\sqrt{\Sigma_r\Sigma_g}).
  \label{equ1}        
\end{align}
where $\mu_r$ and $\Sigma_r$ represent the mean and covariance matrices of the reference set, while $\mu_g$ and $\Sigma_g$ indicate those of synthetic set, and the term \textit{tr} denotes the trace of a matrix, respectively. Thus, Fréchet distance is expected to be 0 if the distribution of the generated samples closely matches that of the real samples, indicating high-quality synthetic samples. Previously, the unbiased FID Infinity method was introduced to address FID's sensitivity to sample size~\cite{chong2020effectively}. Study~\cite{bińkowski2018demystifying} demonstrated that the FID is highly sensitive to experimental settings, which can hinder the reproducibility of results.
\cite{gui2024adapting, tailleur2024correlation} found that the FAD's dependency on embedding choice is closely related to the training data domain, and embeddings from unrelated datasets may limit the FAD's generalizability.
Although the Fréchet distance typically increases with disturbances in synthetic samples~\cite{heusel2017gans, bińkowski2018demystifying}, study~\cite{jayasumana2024rethinking} observed that the FID improved with initial levels of distortion under noise in the VQGAN~\cite{esser2021taming} latent space.

\begin{figure*}[!t]
    \centering
    \begin{subfigure}{.24\linewidth}
      \centering
      \includegraphics[width=\linewidth]{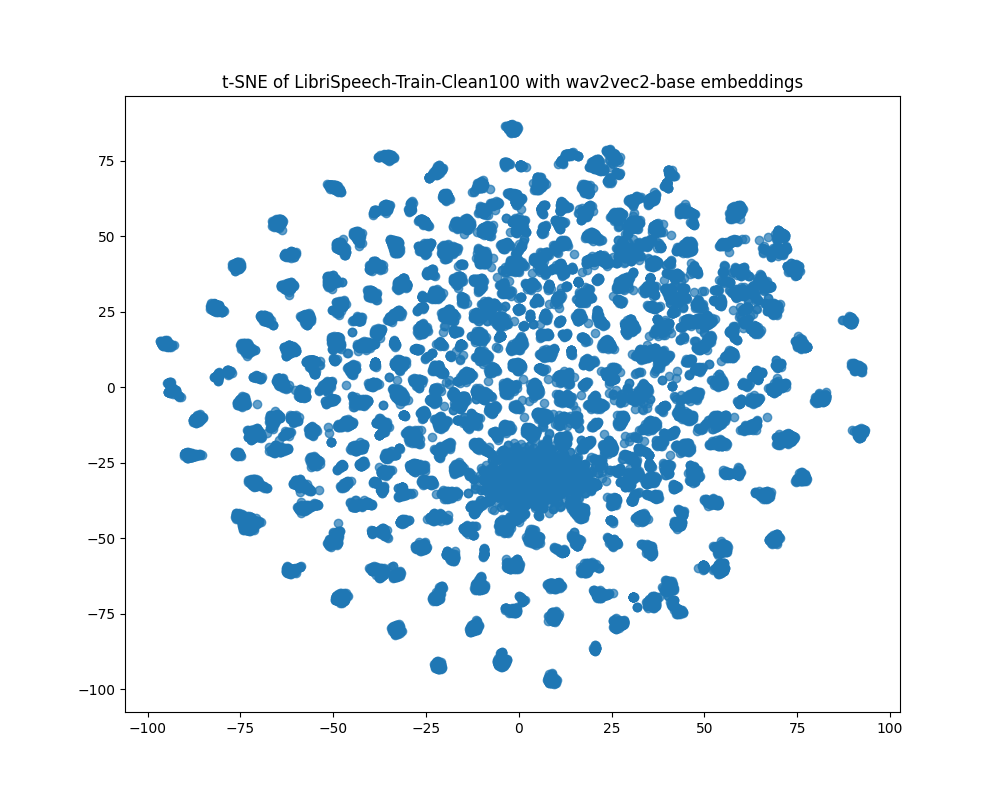}
      \caption{wav2vec2 Base}
      \label{fig:sfig1}
    \end{subfigure}
    \begin{subfigure}{.24\linewidth}
      \centering
      \includegraphics[width=\linewidth]{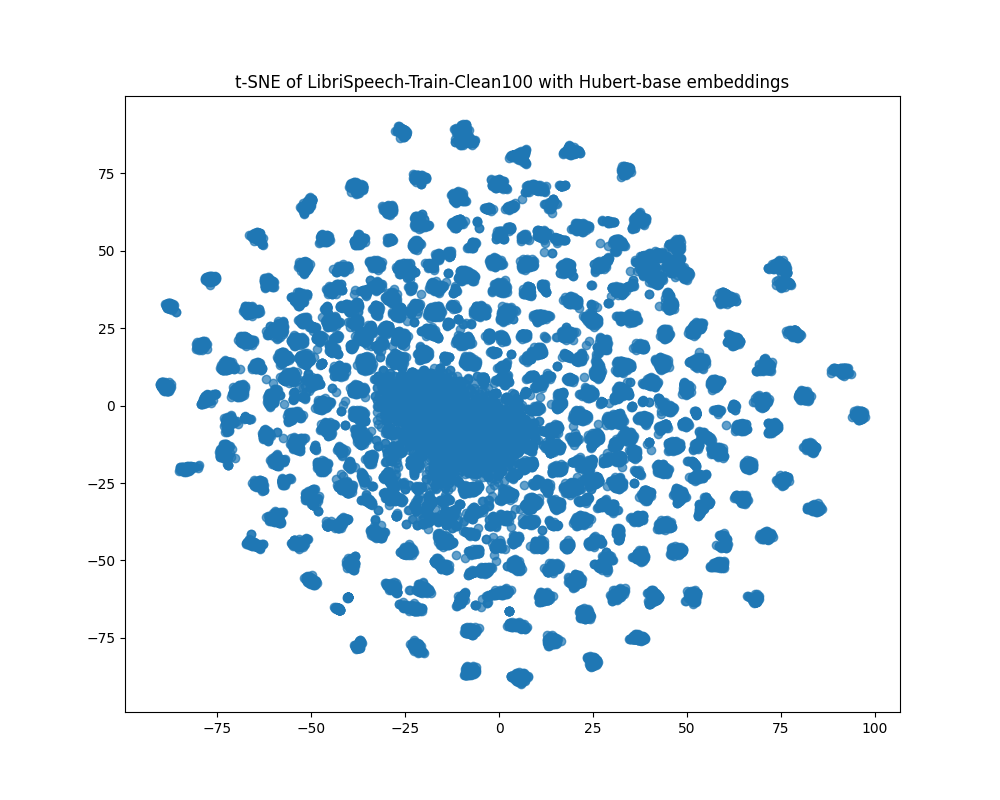}
      \caption{HuBERT Base}
      \label{fig:sfig2}
    \end{subfigure}
    \begin{subfigure}{.24\linewidth}
      \centering
      \includegraphics[width=\linewidth]{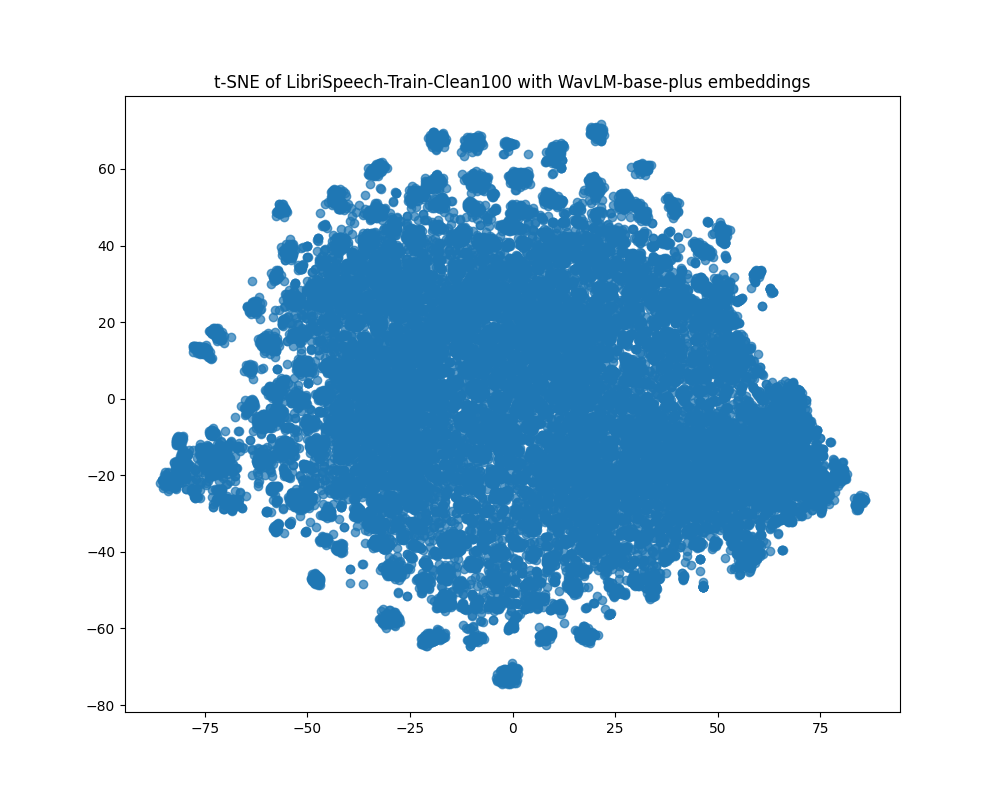}
      \caption{WavLM Base+}
      \label{fig:sfig3}
    \end{subfigure}
    \begin{subfigure}{.24\linewidth}
      \centering
      \includegraphics[width=\linewidth]{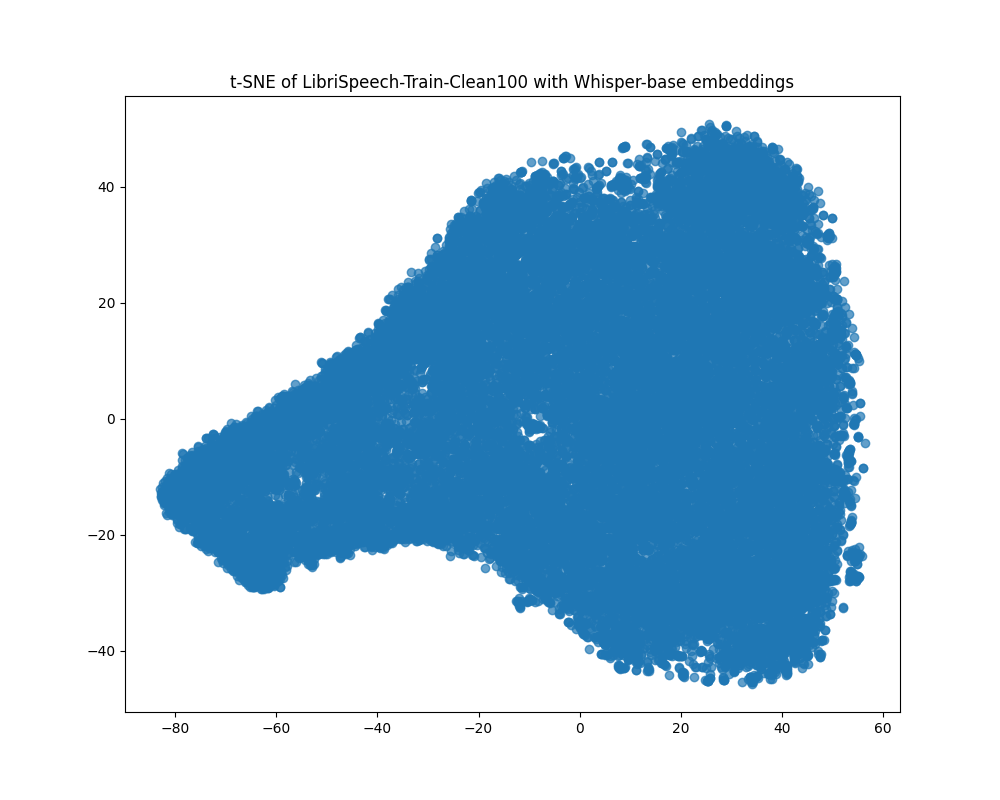}
      \caption{Whisper Base}
      \label{fig:sfig4}
    \end{subfigure}
    \caption{t-SNE results of various speech embeddings of the LS train-clean 100 hour dataset.}
    \label{fig:t-sne}
    \vspace{-5mm}
\end{figure*}

For the TTS, Bińkowski \textit{et al.}~\cite{Bińkowski2020High} computed Fréchet DeepSpeech Distance and Kernel DeepSpeech Distance based on DeepSpeech2~\cite{amodei2016deep} embeddings. VoiceBox~\cite{le2024voicebox} used the Fréchet distance with the 6th layer of a self-supervised wav2vec2.0 embeddings, followed by PCA, while~\cite{he2024emilia} employed pretrained emotion2vec~\cite{ma2023emotion2vec} embeddings. As a result, previous studies in TTS quality evaluation employed varying experimental settings, which lead to different outcomes and complicate users' understanding of the FSD score.

%\vspace{-3mm}

\section{Fréchet Speech Distance}
%Since FSD highly depends on embeddings and datasets for comparison, we first address these issues and throughly analyze that when noise is adding to  real and generated sets, we address these issues and analyze 
\subsection{Speech embeddings and datasets for analyzing FSD}
%\subsubsection{Speech embeddings}
\textbf{Speech embeddings:} Since Fréchet distance relies heavily on the embeddings, we consider employing five distinct speech embeddings for our analysis; self-supervised pretrained wav2vec2 Base~\cite{baevski2020wav2vec}, HuBERT Base~\cite{hsu2021hubert}, WavLM Base+~\cite{chen2022wavlm}, ECAPA-TDNN~\cite{desplanques2020ecapa} and the encoder of Whisper Base~\cite{radford2023robust}. 
wav2vec2 and HuBERT were each pretrained on 960 hours, while ECAPA, WavLM and Whisper were with 2.5K, 94K and 680K hours, respectively.
Our choice of embeddings addresses the potential impact of pretraining data quantity on FID distance calculations~\cite{jayasumana2024rethinking} and shows that out-of-domain embeddings are not useful~\cite{tailleur2024correlation}.
We extracted speech features using each speech embedding, averaged them across output layers, and aggregated them along the time axis. We then calculated mean and covariance matrices from these features.

%\vspace{5mm}
%\subsubsection{Reference and target dataset}
\noindent \textbf{Reference and target dataset:} Choosing datasets for both real and generated samples are also important for FSD measurement. In this work, we use LibriSpeech~\cite{panayotov2015librispeech} (LS) train-clean 100 hour as a reference set due to its sample size of 28K, which is comparable to the COCO~\cite{lin2014microsoft} 30K dataset often used as a reference in text-to-image FID benchmarks. For synthetic speech generation, we utilize LS train-clean 100 hour for acoustic prompts in zero-shot settings, and LS test-clean for text prompts. Consequently, each speaker from the LS train-clean 100 hour dataset will be selected approximately 10 times for synthetic generation in this setting.

\noindent \textbf{Embedding analysis:} To analyze the distribution of points in the reference set, Fig.~\ref{fig:t-sne} presents the 2-dimensional t-SNE visualization of the embeddings, which takes the mean of the features along the time sequences.
As in Fig.~\ref{fig:t-sne}, the t-SNE plots of different embeddings do not seem to be distributed normally, which is a critical assumption for the closed form solution of FD, as also observed in~\cite{jayasumana2024rethinking}. We further evaluate the multivariate normal distribution of four speech embeddings using Mardia's skewness and kurtosis tests~\cite{mardia1970measures}, as well as the Henze-Zirkler~\cite{henze1990class} test. However, none of the embeddings passed the normality test with a $p$ value of 0.0.

\newcommand{\figsize}{\fontsize{12}{14}\selectfont} % Define a custom font size

\begin{figure*}[!t]
    \centering
    \includegraphics[width=0.7\linewidth]{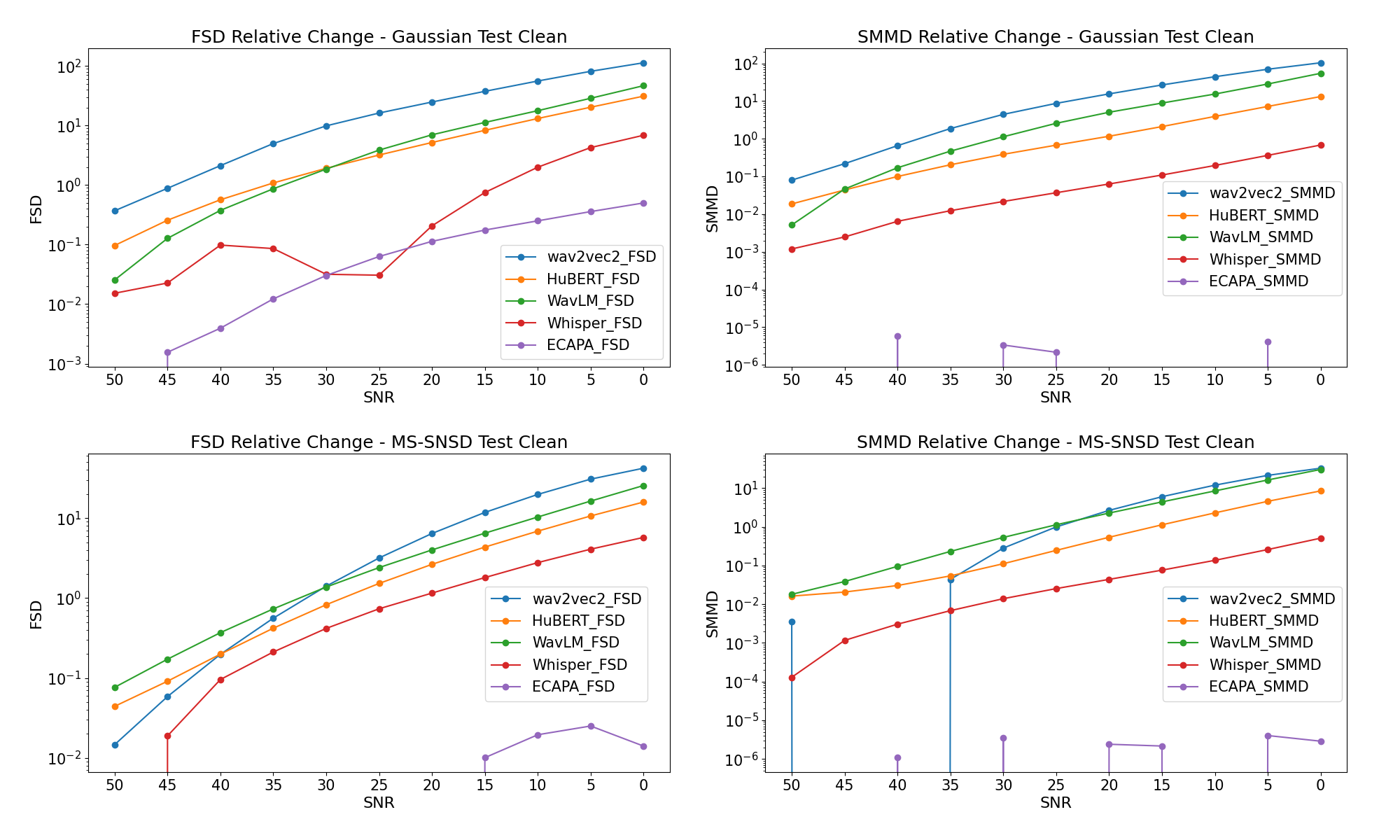}
    \vspace{-5mm}
    \caption{Comparison of the relative change for both FSD and SMMD metrics on a logarithmic scale, applied to two distinct noise sets in the LS test-clean, with SNR values ranging from 0 to 50 dB. Lower SNR values indicate lower sample quality.}
    \label{fig:fig2}
    \vspace{-3mm}
\end{figure*}

\begin{figure*}[!t]
    \centering
    \includegraphics[width=0.65\linewidth]{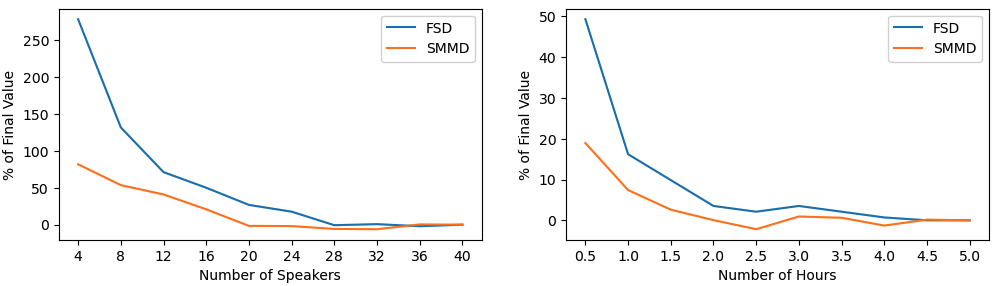}
    \caption{Sample efficiency testing for FSD and SMMD on LS test-clean, using WavLM embeddings. Left: Behavior when subsets selected based on speakers, Right: Behavior for random subset selection.}
    \vspace{-5mm}
    
    \label{fig:fig3}
\end{figure*}

\subsection{Speech Maximum Mean Discrepancy distances}
While FSD provides a strong metric for evaluating synthetic speech quality, prior studies have highlighted its sensitivity to embedding choices, bias issues, and sample distributions~\cite{Bińkowski2020High, bińkowski2018demystifying}. To address this limitation, we also introduce Speech Maximum Mean Discrepancy (SMMD), a Gaussian kernel-based metric that does not rely on the assumption of normally distributed embeddings, thereby providing a complementary perspective in assessing synthetic speech quality.
SMMD can be formularized as follows:
\vspace{-3mm}
\begin{align}
  \mathbf{SMMD}(\mathbf{R}, \mathbf{G}) = \frac{1}{m(m-1)}\sum_{1\leq i,j\leq m}k(\mathbf{R}_{i}, \mathbf{R}_{j})  \nonumber \\[-5pt]
  +\frac{1}{n(n-1)} \sum_{1\leq i, j\leq n}k(\mathbf{G}_i, \mathbf{G}_j) + \sum_{i=1}^m \sum_{j=1}^n k(\mathbf{R}_i, \mathbf{G}_j),
  \label{equ2}
\end{align}
where $i\neq j$, $\mathbf{R}$ and $\mathbf{G}$ are reference and generated sets, $k$ is a kernel function where we use Gaussian kernel as below:
\vspace{-2mm}
\begin{align}
  k(\mathbf{r}, \mathbf{g}) = \exp(-\frac{||r-g||^2}{2\sigma^2})
  \label{equ3}
\end{align}
Here, we use the mean values of features along the time axis.

\subsection{Noise addition for quality evaluation}
To assess noise's effect on FSD's ability to evaluate speech generation quality, we added Gaussian noise with a distribution of $\mathcal{N}(0,1)$ and MS-SNSD noise~\cite{reddy2019scalable} dataset, which contains real-world background noises designed for speech enhancement and denoising tasks, to the LS test-clean dataset at SNR levels ranging from 0 to 50 dB in 5 dB increments.
Including MS-SNSD background noise better simulates real-world complex speech distortions, providing insights into the behavior of both FSD and SMMD in such environments.

Fig.~\ref{fig:fig2} presents the FSD and SMMD scores for various speech embeddings under different noise levels. Generally, FSD and SMMD scores increase with increasing noise levels, except for Whisper Base and ECAPA-TDNN. This suggests that lower FSD and SMMD scores indicate higher quality synthetic samples, similar to the findings in~\cite{le2024voicebox}. However, FSD scores for Whisper in Gaussian noise, especially between 35 and 25 dB, exhibit inconsistent behavior compared to SMMD as noise levels increase. Interestingly, similar trends were observed in~\cite{jayasumana2024rethinking}. %, despite different experimental conditions. 
This may be attributed to Whisper's training objective, which is focused on ASR performance rather than direct feature extraction for speech representations. Unlike self-supervised embeddings, Whisper embeddings may prioritize linguistic content over acoustic characteristics, leading to variations in how noise affects their distribution. %This highlights the importance of selecting embeddings that align with the objective of both FSD and SMMD-based speech quality evaluation.
Moreover, the FSD and SMMD scores of ECAPA do not show the expected trend as noise increases, except in the case of Gaussian test-clean noise, suggesting that ECAPA embeddings are not effective for either metric.
Even though speech embeddings fail the multivariate normality test, FSD generally follows the trends of SMMD, except for Whisper Base. Therefore, selecting suitable speech embeddings is essential for accurate FSD computation. The scale of FSD scores varies markedly between different embeddings and noise settings, which complicates the reproducibility of FSD across studies with different experimental conditions.
%Figure~\ref{fig:fig2} illustrates the FSD and SMMD results according to various speech embeddings with different noise SNR. It can be seen that as noise levels increase, FSD and gSMMD scores generally increase, except for Whisper Base. This suggests that lower FSD and SMMD values are associated with higher-quality synthetic samples, exhibiting a similar phenomenon to what is described in VoiceBox~\cite{le2024voicebox}. However, while SMMD scores gradually increase with stronger noise, the FSD scores for Whisper in Gaussian noise setting, particularly in the 35 to 25 dB range, show inconsistent behavior compared to SMMD.
%Interestingly, despite the differences in experimental settings, similar trends were also reported in~\cite{jayasumana2024rethinking}. 
%Despite speech embeddings not passing the multivariate normality test, our results demonstrate that FSD exhibits the same trends as the unbiased SMMD metric except for Whisper Base.
%Therefore, selecting appropriate speech embeddings is crucial for FSD computation. The scale of FSD scores is significantly different under the two noise settings, which complicates the reproducibility of FSD across studies with different experimental conditions.

\subsection{Sample Efficiency}
Previous works~\cite{chong2020effectively, gui2024adapting, jayasumana2024rethinking} have argued that the Fréchet distance, when applied to both music and image domains, can be biased by sample size and requires a sufficiently large number of samples to converge.
%To assess sensitivity to sample size, 
To address these issues,
we created subsets from the test-clean set by sampling $x\%$ of the utterances, with $x$ ranging from 10 to 100 in increments of 10. We employed two sampling strategies: \textit{random sampling} and \textit{speaker-based sampling}. As shown in Fig.~\ref{fig:fig3}, we observe that FSD behaves similarly to our proposed unbiased SMMD metric and is highly sample-efficient. When randomly sampling, both metrics begin to converge to their final values within 3 hours of speech data. Furthermore, the metrics capture speaker diversity, with sharp increases when the number of speakers in the subset is very small, albeit FSD seems much more sensitive to reduction in speaker diversity.
This sample-efficient behavior makes FSD and SMMD highly practical for real-world applications, such as tracking model performance during TTS model training.

\begin{table*}[t!]
    \centering
    %\vspace{-3mm}
    \caption{Overall synthetic speech quality evaluations with LS test-clean and test-other datasets. Note that TTS intelligibility is measured using both real and synthetic test sets, while synthetic-WER is evaluated only on real audio sets.}
    
    \label{tab:table1}
    \renewcommand{\arraystretch}{1}
    \addtolength{\tabcolsep}{1pt}
    \resizebox{\linewidth}{!}{
    \begin{tabular}{ll|ll|lllll|lllll}
    \toprule
    & &  \multicolumn{2}{c}{ASR}  & \multicolumn{5}{c}{FSD} & \multicolumn{5}{c}{SMMD} \\
    \cmidrule(l{2pt}r{2pt}){3-14}
    & Model & TTS intelligibility & synthetic-WER & wav2vec2 & HuBERT & WavLM & Whisper & ECAPA & wav2vec2 & HuBERT & WavLM & Whisper & ECAPA \\
    \hline \midrule

    \multirow{5}{*}{\rotatebox[origin=c]{90}{\textbf{test-clean}}} & Real Audio & 9.19 & 6.51 & 0.27 & 0.36 & 0.16 & 5.52 & 20407.04 & 1.70 & 2.04 & 0.39 & 38.66 & \multirow{5}{*}{0.42} \\

    & XTTS & 9.97 & 7.81 & 1.41 & 1.57 & 1.06 & 7.71 & 30412.13 & 2.26 & 2.20 & 1.86 & 35.00 &   \\
    & YourTTS & 15.66 & 8.59 &	2.33 &	2.74 &	1.90 &	20.97 &	24635.16 & 4.00 &	4.39 &	2.99 &	55.05 &  \\
    & Tacotron2 &	13.67 &	9.84 &	3.75 &	3.78 &	2.58 &	38.90 & 81563.64 &	4.37 &	4.74 &	3.13 &	45.90 &  \\
    & VITS &	11.91 &	9.03 &	3.86 &	3.86 &	2.44 &	31.87 &	81580.02 & 4.85 &	4.43 &	3.26 &	46.95 &  \\

    \midrule

    \multirow{5}{*}{\rotatebox[origin=c]{90}{\textbf{test-other}}} & Real Audio & 18.38 & 17.83 & 0.96 &	1.05 &	0.60 &	8.89 & 31916.00 &	3.12 &	3.77 &	1.31 &	53.59 & \multirow{5}{*}{0.38} \\

    & XTTS & 11.49 & 20.01 & 1.53 &	1.75 &	1.16 &	9.80 & 31574.66 &	2.78 &	2.73 &	2.22 &	46.95 &  \\
    & YourTTS & 17.59 & 21.51 &	2.51 &	2.98 &	2.05 &	22.40 &	25133.07 &	4.84 &	5.47 &	3.54 &	70.84 &  \\
    & Tacotron2 & 16.67 &	24.85 &	4.28 &	4.11 &	3.17 &	37.82 & 80354.49 &	4.77 &	5.44 &	3.36 &	57.57 &  \\
    & VITS &	12.53 &	23.51 &	3.94 &	3.99 &	2.25 &	29.29 &	80454.38 &	5.34 &	5.06 &	3.48 &	59.39 &  \\

    \bottomrule
    \end{tabular}}
    \vspace{-3mm}
\end{table*}

%\vspace{-2mm}
\section{Synthetic Speech Evaluation Experiments}

\subsection{Settings}
%\subsubsection{TTS Models}
\textbf{TTS Models}: To evaluate the quality of synthetic speech samples, we employ open-source TTS models. For zero-shot multi-speaker settings, we use the XTTS~\cite{casanova2024xtts} and YourTTS~\cite{casanova2022yourtts} models.
We randomly select one of the 251 speakers for the acoustic prompts, resulting in each speaker being used approximately 10 times for speech generation. For single-speaker settings, we utilize the Tacotron2~\cite{shen2018natural} and VITS~\cite{kim2021conditional} models, which rely solely on text prompts. All the model checkpoints in this paper are sourced from Coqui-AI TTS.%\footnote{\url{https://github.com/coqui-ai/TTS}}.

\noindent \textbf{Training details for synthetic-trained ASR}: We fine-tune the Whisper-tiny on synthetic samples for only 10 epochs using the AdamW optimizer, with a learning rate of 5e--5 and a batch size of 16. The synthetic samples denote speech \textit{re-synthesis}, i.e., generated with the same acoustic and text prompts from the LS train-clean 100.

%\vspace{-2mm}
\subsection{Evaluation Metrics}
\vspace{-2mm}
For comprehensive synthetic speech quality evaluations, we employ \textbf{TTS intelligibility} and \textbf{synthetic-trained ASR WER} (synthetic-WER). For TTS intelligibility, we use a Whisper-tiny model to compute WER on both real and synthetic samples, and define \textbf{synthetic-ASR} by fine-tuning this model on synthetic data (LS train-clean 100) and evaluating it on LS test-clean/test-other. 
%For the synthetic-WER, the same model is fine-tuned on synthetic data (LS train-clean 100) and evaluated on LS test-clean/test-other. 
This can estimate the distributional similarity between synthetic and real speech, under the assumption that the closer the WER is to a model trained only on real data, the better the synthetic samples approximate natural~\cite{minixhofer2022evaluating}.
%The synthetic-WER provides an estimate of the distributional similarity between synthetic and real speech, based on the assumption that the closer WER is to an ASR model trained exclusively on real speech, the better the synthetic data approximates natural recordings~\cite{minixhofer2022evaluating}. 

In addition, we also conducted a human listening MOS test to directly assess perceptual quality. Specifically, we sampled 100 utterances by five sources (LibriSpeech reference audio, XTTS, YourTTS, Tacotron2, and VITS). Each audio excerpt was rated for naturalness on a five-point Likert scale. The presentation order of samples was fully randomized across participants to mitigate bias. A total of 32 raters were recruited through Amazon Mechanical Turk, providing a diverse set of subjective evaluations.

%\subsubsection{Training details for synthetic-trained ASR}
 %We removed all symbols for WER.
%To train the Whisper-tiny model with synthetic samples, we use the Hugging Face framework with DeepSpeed\footnote{\url{https://github.com/microsoft/DeepSpeed}} ZeRO configuration 2. The learning rate is set to 5e--6, with a batch size of 8 and accumulation steps per GPU also set to 8. The training is conducted for up to 10 epochs, totaling 600 global steps, with the warmup steps set to 100.

\subsection{Comprehensive Results}
%Comprehensive speech quality evaluations using four different metrics on the LS test-clean and test-other are presented in Table~\ref{tab:table1}. 
Table~\ref{tab:table1} summarizes overall quality evaluations on the LS test-clean and test-other sets. %provides overall speech quality evaluations on the LS test-clean and test-other sets.
\textbf{TTS intelligibility} reveals complex results; zero-shot multi-speaker systems did not consistently have better quality than single-speaker models, always, with Tacotron2 and VITS outperforming YourTTS. %Our findings show that Tacotron2 and VITS outperformed YourTTS. 
Interestingly, synthetic samples exhibit better TTS intelligibility compared to real audio in the LS test-other, likely because they were generated from train-clean speakers and thus easier for AST to recognize~\cite{panayotov2015librispeech}. %. This is because synthetic samples are generated from LS train-clean speakers, making them easier for ASR to recognize~\cite{panayotov2015librispeech}. 
Meanwhile, both FSD and SMMD yielded substantially lower (better) scores for real audio across datasets. These results indicate that while TTS intelligibility is informative for assessing linguistic correctness, \textbf{\emph{it remains limited for capturing overall perceptual quality}}~\cite{le2024voicebox}.
%for real audio are significantly better than those for synthetic audio in both sets. Therefore, while TTS intelligibility can be useful for evaluating synthesized samples from a linguistic perspective (content correctness), it appears to have limitations in assessing the overall quality of synthesized speech, as described in~\cite{le2024voicebox}.

For the \textbf{synthetic-WER} on the test-clean set, all results improved when fine-tuned on their respective synthetic data. Although Tacotron2 and VITS showed higher intelligibility than YourTTS, multi-speaker models achieved lower synthetic-WER, suggesting better ASR generalization from diverse speaker data.
%Notably, while Tacotron2 and VITS achieved higher TTS intelligibility than yourTTS, multi-speaker TTS models surpassed single-speaker models in synthetic-WER. This may be attributed to the enhanced ASR generalizability by fine-tuning with multi-speaker data compared to single-speaker training. 
In contrast, all results degraded on test-other, indicating that LS train-100 was insufficient for cross-condition adaptation.
%Unlike test-clean, all synthetic results in test-other showed degradation after fine-tuning, i.e., LS train 100 proved insufficient for enhancing LS test-other performance. 
Using the full LS train-960 yielded substantial gains in our preliminary study (5.03 and 14.16 on test-clean and test-other, respectively). 
%We observed that fine-tuning with the full LS train 960 hours led to improvements of 5.03 and 14.16 in synthetic-WER on test-clean and test-other, respectively. 
Importantly, \emph{synthetic-WER trends were consistent with FSD}, supporting \emph{\textbf{FSD as a cost-efficient alternative to ASR-based evaluations}}.
%\emph{Synthetic-WER results strongly align with the FSD}, indicating that \emph{\textbf{FSD could be a cost-effective substitute for ASR training based approaches.}}

Our findings indicate that \textbf{SMMD}'s effectiveness is highly dependent on the choice of embedding. SMMD of Whisper showed inconsistent trends compared to FSD and synthetic-WER, and ECAPA-TDNN produced nearly identical values across systems, failing to capture differences. %that the real audio set performs worse than XTTS, whereas FSD and synthetic-WER exhibited opposite trends. Based on ECAPA-TDNN, SMMD reported very similar values for all audios. 
SMMD also did not reflect speaker diversity on test-other, whereas \emph{\textbf{FSD on WavLM}} successfully distinguished these variations. %failed to capture speaker diversity on test-other; despite significant differences in synthetic-WER, except for XTTS, SMMD of WavLM showed less variation. However, \emph{\textbf{FSD on WavLM} better captured these differences.} 
Moreover, WavLM embeddings provided the most stable and consistent results in both FSD and SMMD, yielding the lowest distance values for real audio.
%Moreover, while the ordering of FSD results for wav2vec2 and HuBERT differed from those of SMMD, when comparing to synthetic-WER (The TTS intelligibility scores for Tacotron2 and VITS are 16.37 and 11.91, respectively, whereas their SMMD scores are 4.376 and 4.846), only WavLM demonstrated stable results. Also, WavLM embeddings showed the lowest absolute values in both FSD and SMMD, for real audio evaluations, as desired. 
Overall, our results suggest \textit{a positive correlation between \textbf{synthetic-WER and both FSD and SMMD}}, with systems exhibiting lower (better) synthetic-WER also showing lower (better) FSD and SMMD values. 

\subsection{Correlation between Human and Objective Metrics}

\begin{table}[!t]
\centering
\caption{Human MOS scores and objective metric values (FSD, SMMD) for reference and each TTS-generated sample.}
%\vspace{-3mm}
\renewcommand{\arraystretch}{1}
    \addtolength{\tabcolsep}{1pt}
    \resizebox{\linewidth}{!}{
\begin{tabular}{llcccc}
\hline
\textbf{System} & \textbf{Type} & \textbf{MOS} & \textbf{FSD (WavLM)} & \textbf{SMMD (WavLM)} \\
\hline
Real      & Reference & 4.52 ± 0.13  & 0.16  & 2.04  \\
XTTS      & Multi-speaker & 4.16 ± 0.43  & 1.06  & 2.20  \\
YourTTS   & Multi-speaker & 3.75 ± 0.28  & 1.90  & 4.39  \\
Tacotron2 & Single-speaker & 3.63 ± 0.22  & 2.58  & 4.74  \\
VITS      & Single-speaker & 4.25 ± 0.11 & 2.44  & 4.43  \\
\hline
\end{tabular}}
\label{tab:mos_metrics}
\vspace{-5mm}
\end{table}
%To further investigate the relationship between human-perceived naturalness and objective metrics, we examined the correlation between the UTMOS scores and both the FSD and SMMD across the evaluated TTS systems, as in Table~\ref{tab:mos_metrics}. We observed that systems with lower (better) FSD and SMMD scores generally aligned with higher MOS ratings. However, the alignment was not perfect. For example, while WavLM-based FSD and SMMD captured relative differences effectively, some inconsistencies were observed, particularly for Whisper and ECAPA embeddings, where SMMD did not reflect MOS trends as accurately. These findings highlight that while objective metrics like FSD and SMMD provide useful insights into synthetic speech quality, they may not fully account for perceptual aspects captured by human listeners, especially when system design factors (e.g., single-speaker vs. multi-speaker) are involved. Overall, embedding choice critically impacts the validity of objective metrics relative to human judgment.
To validate the perceptual relevance of objective metrics, we compared the human MOS ratings with the corresponding FSD and SMMD scores for each system, as in Table~\ref{tab:mos_metrics}. 
Overall, we observed that systems with lower (better) FSD and SMMD scores generally aligned with higher MOS ratings. Interestingly, however, the correspondence was not perfect. VITS achieved a relatively high MOS (4.25), close to XTTS, despite having worse FSD and SMMD scores than XTTS. This suggests that single-speaker systems can yield perceptually natural speech even when distance-based metrics indicate distributional divergence from the reference data. In contrast, multi-speaker model such as XTTS obtained strong MOS scores alongise comparatively lower distance metrics, highlighting the trade-off between speaker diversity and distributional similarity. 

These results indicate that while FSD and SMMD provide cost-efficient and reproducible approximations of subjective quality, they \textbf{\emph{cannot fully replace human evaluation. Instead, they should be interpreted as complementary measures.}}
Importantly, such objective metrics become particularly valuable in scenarios where large-scale data must be evaluated or where direct listening-based assessments are infeasible, offering a practical alternative for tacking relative quality across systems.

%Overall, systems with higher MOS values tended to exhibit lower (better) FSD and SMMD scores, indicating that both metrics capture important aspects of human-perceived quality.

%\vspace{-3mm}
%Although SMMD is well-regarded as an unbiased estimator~\cite{bińkowski2018demystifying, Bińkowski2020High}, our findings indicate that its effectiveness is highly dependent on the choice of embedding. For example, the gSMMD of Whisper showed that the ground truth performs worse than XTTS, while FSD and synthetic trained ASR WER exhibited different trends. Additionally, while the FSD scores for wav2vec2 and HuBERT differed from those of gSMMD, only WavLM demonstrated stable results. Notably, WavLM also showed the worst performance in Tacotron2 in terms of FSD and SMMD. As a result, there seems to be a positive correlation between synthetic trained ASR WER and FSD as well as SMMD. Systems with higher synthetic trained ASR WER tend to have higher FSD values, %which indicates that the more errors a TTS system makes (low quality), the more it deviates from the ground truth in terms of feature space (higher FSD). 
%Therefore, we recommend using WavLM for stable FSD performance, as it provides accurate evaluations of synthetic speech quality.

\section{Conclusion}

%This study 
We comprehensively evaluated several speech embeddings for FSD and SMMD under varying noise conditions and sample sizes, and compared them against objective metrics and human MOS. %such as TTS intelligibility and synthetic WER. 
Our findings indicate that the choice of embedding is critical for reliable evaluation of synthetic speech quality. %an accurate evaluation of the quality of the synthetic speech sample. 
%In particular, WavLM Base+ features provided the most consistent alignment with objective measures. 
While FSD and SMMD cannot fully replace human listening tests, they serve as complementary tools that are cost-efficient, reproducible, and valuable when large-scale evaluations or direct perceptual assessments are impractical.
%The source code used in this study is available at \url{https://github.com/}.
%We further demonstrated that WavLM Base+ features capture audio more effectively for FSD, compared to SMMD and other embeddings. %Therefore, we recommend using WavLM Base+ features for FSD to assess the quality of speech samples.

\newpage

%\section*{Acknowledgemen t}
%This research was supported by Brian Impact Foundation, a non-profit organization dedicated to the advancement of science and technology for all.

\bibliographystyle{IEEEbib}
\small
\bibliography{refs}

@string{icassp = "Proc. ICASSP"}

@string{interspeech = "Proc. Interspeech"}

@string{icml = "Proc. ICML"}

@string{iclr = "Proc. ICLR"}

@string{eccv = "Proc. ECCV"}

@string{cvpr = "Proc. CVPR"}

@article{casanova2024xtts,
  title={XTTS: a Massively Multilingual Zero-Shot Text-to-Speech Model},
  author={Casanova, Edresson and Davis, Kelly and G{\"o}lge, Eren and G{\"o}knar, G{\"o}rkem and Gulea, Iulian and Hart, Logan and Aljafari, Aya and Meyer, Joshua and Morais, Reuben and Olayemi, Samuel and others},
  journal={arXiv preprint arXiv:2406.04904},
  year={2024}
}

@article{le2024voicebox,
  title={Voicebox: Text-guided multilingual universal speech generation at scale},
  author={Le, Matthew and Vyas, Apoorv and Shi, Bowen and Karrer, Brian and Sari, Leda and Moritz, Rashel and Williamson, Mary and Manohar, Vimal and Adi, Yossi and Mahadeokar, Jay and others},
  journal={Advances in neural information processing systems},
  volume={36},
  year={2024}
}

@inproceedings{casanova2022yourtts,
  title={Yourtts: Towards zero-shot multi-speaker tts and zero-shot voice conversion for everyone},
  author={Casanova, Edresson and Weber, Julian and Shulby, Christopher D and Junior, Arnaldo Candido and G{\"o}lge, Eren and Ponti, Moacir A},
  booktitle={Proc. ICML},
  year={2022}
}

@article{streijl2016mean,
  title={Mean opinion score (MOS) revisited: methods and applications, limitations and alternatives},
  author={Streijl, Robert C and Winkler, Stefan and Hands, David S},
  journal={Multimedia Systems},
  volume={22},
  number={2},
  pages={213--227},
  year={2016},
  publisher={Springer}
}

@article{dowson1982frechet,
  title={The Fr{\'e}chet distance between multivariate normal distributions},
  author={Dowson, DC and Landau, BV666017},
  journal={Journal of multivariate analysis},
  volume={12},
  number={3},
  pages={450--455},
  year={1982},
  publisher={Elsevier}
}

@article{heusel2017gans,
  title={Gans trained by a two time-scale update rule converge to a local nash equilibrium},
  author={Heusel, Martin and Ramsauer, Hubert and Unterthiner, Thomas and Nessler, Bernhard and Hochreiter, Sepp},
  journal={Advances in neural information processing systems},
  year={2017}
}

@inproceedings{48813,title	= {Fr\'echet Audio Distance: A Reference-free Metric for Evaluating Music Enhancement Algorithms},author	= {Dominik Roblek and Kevin Kilgour and Matt Sharifi and Mauricio Zuluaga}, booktitle	= {Proc. Interspeech},
  year={2019}}

@inproceedings{gui2024adapting,
  title={Adapting frechet audio distance for generative music evaluation},
  author={Gui, Azalea and Gamper, Hannes and Braun, Sebastian and Emmanouilidou, Dimitra},
booktitle={Proc. ICASSP},
  year={2024}
}

@article{tailleur2024correlation,
  title={Correlation of Fr$\backslash$'echet Audio Distance With Human Perception of Environmental Audio Is Embedding Dependant},
  author={Tailleur, Modan and Lee, Junwon and Lagrange, Mathieu and Choi, Keunwoo and Heller, Laurie M and Imoto, Keisuke and Okamoto, Yuki},
  journal={arXiv preprint arXiv:2403.17508},
  year={2024}
}

@inproceedings{
Bińkowski2020High,
title={High Fidelity Speech Synthesis with Adversarial Networks},
author={Mikołaj Bińkowski and Jeff Donahue and Sander Dieleman and Aidan Clark and Erich Elsen and Norman Casagrande and Luis C. Cobo and Karen Simonyan},
booktitle={Proc. ICLR},
  year={2020}
}

@inproceedings{chong2020effectively,
  title={Effectively unbiased fid and inception score and where to find them},
  author={Chong, Min Jin and Forsyth, David},
booktitle={Proc. CVPR},
  year={2020}
}

@inproceedings{
bińkowski2018demystifying,
title={Demystifying {MMD} {GAN}s},
author={Mikołaj Bińkowski and Dougal J. Sutherland and Michael Arbel and Arthur Gretton},
booktitle={Proc. ICLR},
  year={2018}
}

@inproceedings{amodei2016deep,
  title={Deep speech 2: End-to-end speech recognition in english and mandarin},
  author={Amodei, Dario and Ananthanarayanan, Sundaram and Anubhai, Rishita and Bai, Jingliang and Battenberg, Eric and Case, Carl and Casper, Jared and Catanzaro, Bryan and Cheng, Qiang and Chen, Guoliang and others},
booktitle={Proc. ICML},
  year={2016}
}

@inproceedings{jayasumana2024rethinking,
  title={Rethinking fid: Towards a better evaluation metric for image generation},
  author={Jayasumana, Sadeep and Ramalingam, Srikumar and Veit, Andreas and Glasner, Daniel and Chakrabarti, Ayan and Kumar, Sanjiv},
booktitle={Proc. CVPR},
  year={2024}
}

@article{baevski2020wav2vec,
  title={wav2vec 2.0: A framework for self-supervised learning of speech representations},
  author={Baevski, Alexei and Zhou, Yuhao and Mohamed, Abdelrahman and Auli, Michael},
  journal={Advances in neural information processing systems},
  year={2020}
}

@article{hsu2021hubert,
  title={Hubert: Self-supervised speech representation learning by masked prediction of hidden units},
  author={Hsu, Wei-Ning and Bolte, Benjamin and Tsai, Yao-Hung Hubert and Lakhotia, Kushal and Salakhutdinov, Ruslan and Mohamed, Abdelrahman},
  journal={IEEE/ACM transactions on audio, speech, and language processing},
  volume={29},
  pages={3451--3460},
  year={2021},
  publisher={IEEE}
}

@article{chen2022wavlm,
  title={Wavlm: Large-scale self-supervised pre-training for full stack speech processing},
  author={Chen, Sanyuan and Wang, Chengyi and Chen, Zhengyang and Wu, Yu and Liu, Shujie and Chen, Zhuo and Li, Jinyu and Kanda, Naoyuki and Yoshioka, Takuya and Xiao, Xiong and others},
  journal={IEEE Journal of Selected Topics in Signal Processing},
  volume={16},
  number={6},
  pages={1505--1518},
  year={2022},
  publisher={IEEE}
}

@inproceedings{radford2023robust,
  title={Robust speech recognition via large-scale weak supervision},
  author={Radford, Alec and Kim, Jong Wook and Xu, Tao and Brockman, Greg and McLeavey, Christine and Sutskever, Ilya},
booktitle={Proc. ICML},
year={2024}
}

@article{he2024emilia,
  title={Emilia: An Extensive, Multilingual, and Diverse Speech Dataset for Large-Scale Speech Generation},
  author={He, Haorui and Shang, Zengqiang and Wang, Chaoren and Li, Xuyuan and Gu, Yicheng and Hua, Hua and Liu, Liwei and Yang, Chen and Li, Jiaqi and Shi, Peiyang and others},
  journal={arXiv preprint arXiv:2407.05361},
  year={2024}
}

@article{ma2023emotion2vec,
  title={emotion2vec: Self-supervised pre-training for speech emotion representation},
  author={Ma, Ziyang and Zheng, Zhisheng and Ye, Jiaxin and Li, Jinchao and Gao, Zhifu and Zhang, Shiliang and Chen, Xie},
  journal={arXiv preprint arXiv:2312.15185},
  year={2023}
}

@inproceedings{reddy2019scalable,
  title={A Scalable Noisy Speech Dataset and Online Subjective Test Framework},
  author={Reddy, Chandan KA and Beyrami, Ebrahim and Pool, Jamie and Cutler, Ross and Srinivasan, Sriram and Gehrke, Johannes},
  journal={Proc. Interspeech},
    year={2019}
}

@inproceedings{panayotov2015librispeech,
  title={Librispeech: an asr corpus based on public domain audio books},
  author={Panayotov, Vassil and Chen, Guoguo and Povey, Daniel and Khudanpur, Sanjeev},
booktitle={Proc. ICASSP},
year={2015}
}

@inproceedings{lin2014microsoft,
  title={Microsoft coco: Common objects in context},
  author={Lin, Tsung-Yi and Maire, Michael and Belongie, Serge and Hays, James and Perona, Pietro and Ramanan, Deva and Doll{\'a}r, Piotr and Zitnick, C Lawrence},
booktitle={Proc. ECCV},
year={2014}
}

@inproceedings{shen2018natural,
  title={Natural tts synthesis by conditioning wavenet on mel spectrogram predictions},
  author={Shen, Jonathan and Pang, Ruoming and Weiss, Ron J and Schuster, Mike and Jaitly, Navdeep and Yang, Zongheng and Chen, Zhifeng and Zhang, Yu and Wang, Yuxuan and Skerrv-Ryan, Rj and others},
booktitle={Proc. ICASSP},
year={2018}
}

@inproceedings{kim2021conditional,
  title={Conditional variational autoencoder with adversarial learning for end-to-end text-to-speech},
  author={Kim, Jaehyeon and Kong, Jungil and Son, Juhee},
booktitle={Proc. ICML},
year={2021}
}

@article{mardia1970measures,
  title={Measures of multivariate skewness and kurtosis with applications},
  author={Mardia, Kanti V},
  journal={Biometrika},
  volume={57},
  number={3},
  pages={519--530},
  year={1970},
  publisher={Oxford University Press}
}

@article{henze1990class,
  title={A class of invariant consistent tests for multivariate normality},
  author={Henze, Norbert and Zirkler, Bernd},
  journal={Communications in statistics-Theory and Methods},
  volume={19},
  number={10},
  pages={3595--3617},
  year={1990},
  publisher={Taylor \& Francis}
}

@article{desplanques2020ecapa,
  title={Ecapa-tdnn: Emphasized channel attention, propagation and aggregation in tdnn based speaker verification},
  author={Desplanques, Brecht and Thienpondt, Jenthe and Demuynck, Kris},
  journal={arXiv preprint arXiv:2005.07143},
  year={2020}
}

@article{minixhofer2022evaluating,
  title={Evaluating and reducing the distance between synthetic and real speech distributions},
  author={Minixhofer, Christoph and Klejch, Ond{\v{r}}ej and Bell, Peter},
  journal={arXiv preprint arXiv:2211.16049},
  year={2022}
}

@inproceedings{esser2021taming,
  title={Taming transformers for high-resolution image synthesis},
  author={Esser, Patrick and Rombach, Robin and Ommer, Bjorn},
  booktitle={Proceedings of the IEEE/CVF conference on computer vision and pattern recognition},
  pages={12873--12883},
  year={2021}
}

\end{document}